\documentclass[aps,prb,twocolumn,nofootinbib, superscriptaddress,10pt,floatfix]{revtex4-2}
\usepackage{amsmath,amsthm,amsfonts,amssymb,amscd,bbold,mathrsfs,bm}
\usepackage[T1]{fontenc}
\usepackage{dcolumn}
\usepackage{physics}
\usepackage{graphicx}
\usepackage{epstopdf}
\usepackage{pgf}
\usepackage{tabularx}
\usepackage{subfigure}
\usepackage{float}
\usepackage{hyperref}

\begin{document}

\title{Toward axion signal extraction in semiconductor spin qubits via spectral engineering}

\author{Xiangjun Tan}
\email{xinshijietxj@gmail.com}
\affiliation{Center on Frontiers of Computing Studies, Peking University, Beijing 100871, China}
\author{Zhanning Wang}
\affiliation{School of Physics, University of New South Wales, Sydney, NSW 2052, Australia}
\date{\today}

\begin{abstract}
Recent advances in quantum sensing and computational technologies indicate the possibility of improving the precision of measurements aimed at detecting cosmological particles and weakly interacting massive particles using various qubit platforms.
While recent progress has been made, mitigating environmental noise remains a challenge in extracting particle parameters with high fidelity.
Addressing these challenges requires efforts on two levels.
At the device level, the qubit and its array acting as a probe must be isolated from electrical and magnetic noise through optimized device geometry.
At the signal processing level, it is necessary to develop filtering methods targeting specific noise spectra based on different qubit architectures.
In this work, we explore the possibility of using semiconductor quantum dot spin qubits as a platform to search for quantum chromodynamics (QCD) axions and, more broadly, axion-like particles.
Starting by deriving an effective Hamiltonian for electron-axion interactions, we identify an axion-induced effective magnetic field and determine the characteristic axion oscillation frequency.
To suppress charge noise in the devices and environmental noise, we first analyze the charge noise spectrum and then develop a dedicated filtering and noise-reduction protocol, paving the way for exploring feasible axion mass ranges.
Our preliminary study holds promise for enhancing the screening of various axion signals using quantum technologies.
We expect that our analysis and filtering protocol can help advance the use of semiconductor quantum dot spin qubit arrays in axion detection.
\end{abstract}
\maketitle

\section{Introduction}
\label{S1: Introduction}

The first model for the axion field, established in 1977, provides a new symmetry mechanism that solves the strong CP problem in quantum chromodynamics \cite{Peccei1977, Weinberg1978, Wilczek1982}.
However, the weak coupling between axions and other Standard Model particles makes the direct experimental detection of axion-related signals difficult.
Moreover, the signal is often overwhelmed by environmental and measurement noise.
Earlier detection proposals considered axion–electromagnetic field couplings, demonstrating that axion fields can be incorporated into Maxwell’s equations by adding extra source terms.
This reformulation transforms elusive axion fields into detectable electromagnetic signals \cite{Sikivie1983}.
Building on this idea, various experimental and theoretical studies over the past decades have significantly improved the constraints on axion parameter boundaries, including the axion decay constant ($f_a$), the axion–photon coupling constant ($g_{a\gamma\gamma}$), and the axion–electron coupling constant ($g_{ae}$) \cite{Duffy2009, Du2018, Posen2023}.
A typical experimental architecture relies on a resonant microwave cavity within a strong superconducting magnet to detect axion–photon conversion at cryogenic temperatures \cite{Asztalos2010, Semertzidis2022, Roberto2022,Sushkov2023}.
Pioneering works, such as the transmon‐based Qub‐IT approach, use a high‐$Q$ superconducting cavity operating in GHz range, combined with quantum non-demolition single‐photon counting via fixed‐frequency transmon qubits operated below 20~mK \cite{Moretti2024}.
While promising, this method requires a precise design of the cavity resonant frequency, which may limit the frequency bandwidth for axion searches.

Meanwhile, the rapidly expanding field of quantum computing—particularly on semiconductor quantum-dot platforms—offers new opportunities for axion detection, thanks to its scalability enabled by CMOS compatibility and its compact layout, which allows for qubit array operations simpler \cite{Chan2018, GonzalezZalba2021, Wang2022}.
Electron spin qubits in semiconductor quantum dots encode quantum information in their spin, forming a natural two-level system under an external magnetic field as described by the Zeeman Hamiltonian \cite{Loss1998}.
Advances in semiconductor processing have improved quantum dot quality across multiple material platforms, particularly III–V semiconductors (e.g. GaAs, InAs) and group-IV materials (e.g. Si, Ge) \cite{Burkard2023}.
Group IV semiconductors offer two further advantages.
First, they can be isotopically purified to minimize hyperfine contact interactions, thereby improving qubit dephasing times \cite{Wang2021}.
Second, their non-polar crystal structure suppresses piezoelectric phonon coupling, leading to longer spin-relaxation times and enabling coherent Rabi oscillations \cite{Sarkar2023}.
Additionally, the long coherence times of electron spin qubits open up possibilities for multi-qubit entanglement, including the preparation of squeezed states, Bell states, and Greenberger–Horne–Zeilinger (GHZ) states, thereby enhancing quantum-sensing sensitivity \cite{Eldredge2018, Steinacker2025}.
Over the past two decades, spin qubits have shown promise for quantum sensing and metrology applications \cite{Luca2018}.
In particular, theoretical proposals suggest that arrays of quantum dot spin qubits could serve as platforms to search for weakly interacting massive particles and to test objective-collapse models such as the Diósi–Penrose mechanism, offering a potential probe of physics beyond the Standard Model \cite{Bose2025}.

Recent experiments have also demonstrated industry-level fabrication of ultra-fast, coherently operating spin qubits, including fabrication optimizations, coherent spin state control via spin resonance methods (such as electron spin resonance and electric dipole spin resonance), and reliable qubit state readout \cite{Scappucci2021,Elsayed2024}.
This precise control over spin qubits enables the potential of searching for fine energy shifts due to variations in the magnetic field induced by axions, providing a platform to probe the effective magnetic field generated by axions.
Although axions couple to electron spins with extremely small coupling constants, this interaction can modulate the effective $g$-factor, which is reflected in changes to the qubit Larmor frequency.
Therefore, by taking advantage of the precise phase and amplitude measurements, and employing a CMOS-based $N$-qubit system coupled to a high-$Q$ resonator, it becomes possible to scan axion frequency windows and search for signals induced by axions \cite{GonzalezZalba2021, Chen2024, Eggli2025}.

Another major challenge for axion detection is mitigating environmental and device noise \cite{Yu2022, Lu2015}.
For a single qubit at millikelvin temperatures, electrical noise from single-charge defects and from ensemble $1/f$ fluctuations can be suppressed by optimizing magnetic-field orientation and implementing strain engineering \cite{Culcer2013,Bermeister2014,Martinez2024,Wang2025}.
Additionally, the noise spectrum of electron spin qubits can be well characterized, allowing us to filter out unwanted noise by targeting specific frequency ranges, thereby improving the signal-to-noise ratio (SNR) at the outset \cite{Galperin2006, Connors2019}.
Subsequently, a series of post-processing noise-filtering techniques can be implemented to further enhance signal quality, which forms the central aim of our work.

In this manuscript, we first introduce the axion model, followed by the Hamiltonian of the electron spin qubit platform.
To relate the general axion framework to a concrete realization, we adopt the Dine–Fischler–Srednicki–Zhitnitsky (DFSZ) model as our specific axion candidate \cite{Zhitnitskii1980, Dine1981}.
The DFSZ axion model arises from a two-Higgs-doublet extension of the Standard Model, in which the Peccei–Quinn symmetry is carried by both quarks and leptons via Yukawa interactions.
The tree‐level axion–electron coupling is given by: 
\begin{equation}
    \mathcal{L}_{ae} = -i g_{ae} a \bar{\psi}\gamma_5\psi, \quad
    g_{ae} = C_e \frac{m_e}{f_a} \,,
\end{equation}
where $f_a$ is the axion decay constant, and $m_e$ is the electron mass.
The coupling coefficient $C_e$ is:
\begin{equation}
    C_e = \frac{\cos^2\beta_a}{3} \, \text{DFSZ-I}, 
    \quad C_e = \frac{\sin^2\beta_a}{3} \, (\text{DFSZ-II})\,,
\end{equation}
with $\beta_a$ being the Higgs vacuum alignment angle.
Our detection scheme relies on direct coupling of the axion field gradient $\nabla a$ to the electron spin polarization.
The DFSZ model provides a direct, non-zero axion-electron coupling ($g_{ae}$) at tree level, in contrast to the Kim–Shifman–Vainshtein–Zakharov (KSVZ) model, where such coupling arises only indirectly through heavy quark \cite{Shifman1980, Kim1985}.
Therefore, the DFSZ maximizes the conversion of the axion-induced effective pseudo-magnetic field into spin precession, which can be smoothly integrated into the spin-qubit platform.
Table~\ref{tab1} below summarizes the parameters corresponding estimated bounds of the DFSZ model \cite{Duffy2009, Yang2024}.
\begin{table}
\begin{ruledtabular}
\begin{tabular}{lll}
Parameter & Range & Remarks \\
\hline
$f_a$ & $10^9$--$10^{12}$ & decay constant (GeV) \\
$m_a$ & $\leq 10^{-3}$ & axion mass (eV) \\
$g_{a\gamma\gamma}$ & $\sim 10^{-14}$ & axion--photon coupling (GeV$^{-1}$) \\
$g_{ae}$ & $\leq 10^{-13}$ & axion--electron coupling (dimensionless) \\
$\rho_a$ & $0.3$ & local dark matter density (GeV cm$^{-3}$) \\
\end{tabular}
\end{ruledtabular}
\caption{DFSZ model parameters and estimated bounds. 
Due to the wide variety of axion models, the parameter ranges listed here are those specifically relevant to our work.}
\label{tab1}
\end{table}

Then, we can analyze the qubit–axion interactions in the non-relativistic limit ($\dot{a}^2 \gg (\nabla a)^2$) and derive an effective qubit Hamiltonian.
From this effective Hamiltonian, we extract the energy spectrum variations due to the magnetic field induced by the axion field, and use standard density-matrix (Lindblad) equations to compute the spin-polarization dynamics, which is the experimentally measurable quantity.
For the postprocessing, this subtly modulated interaction signal is passed through a band‐pass filter centered on the expected sideband frequencies to constrain the analysis bandwidth and reject out‐of‐band noise. While this filtering does not reduce the fundamental noise spectral density at the sidebands—and thus does not alter the intrinsic frequency‐domain SNR—it confines noise power to a narrower band, making the two modulation sidebands more readily distinguishable against a suppressed out‐of‐band noise background.

The main conclusions of our work are as follows:
1.) In the non-relativistic limit, by treating the axion field gradient $\nabla a$ as an effective magnetic field in the qubit Hamiltonian, we derive a  corresponding modulation of the qubit Larmor frequency and spin rotation.
2.) We design a multi-stage filtering and bandwidth-engineering schemes based on next-generation CMOS qubit array and a high-$Q$ resonator to improve the SNR.

\section{Methodology}
\label{S2: Methodology}

In this section, we first review the model of semiconductor quantum dot-based electron spin qubits and the axion model as a pseudo-scalar Nambu–Goldstone particle.
Next, we discuss the effective axion–electron Hamiltonian within the framework of axion electrodynamics.
By analyzing spin-polarization dynamics, we demonstrate how axion–qubit interactions produce a measurable modulation in the total polarization signal.
Finally, we introduce the minimal blueprint for a possible experimental setup to detect the axion signals in the semiconductor quantum dot spin qubit platform, and then present our signal-filtering and amplification schemes.
In the following discussion, we use the nature units.

\subsection{Electron spin qubit}
\label{S2: SS1 Electron spin qubit}
We consider an electron confined in a silicon semiconductor quantum dot.
Here, we focus on a single valley in silicon by assuming a large valley splitting, which can be achieved via strain engineering.
In general, the valley degrees of freedom may introduce additional relaxation mechanisms, which which are beyond the scope of this work \cite{Culcer2010}.
Firstly, a two-dimensional electron gas is formed at the Si/SiO${}_2$ interface by a gate field along z-direction \cite{Pla2012}.
Within the quantum dot plane, the confinement potential can be described as a two-dimensional harmonic oscillator potential denoted as $V(x,y)=(x^2+y^2)/(2 m^* a_0^4)$.
Accordingly, the electron wavefunction in the $xy$-plane can be solved by the Fock-Darwin states \cite{Bermeister2014}.
The quantum dot size $a_0$ is set to be 60~nm, and $m^*=0.19m_0$ is the effective mass in silicon.
The total Hamiltonian, in the quasi-two-dimensional limit, for an electron in a silicon quantum dot reads:
\begin{equation}
  \label{Eq: QD total Hamiltonian}
  H = \frac{k^2}{2m^*} + V(x,y) + H_{\text{SO}} + H_{\text{Z}} \,,
\end{equation}
where $m^*$ denotes the effective mass.
$H_{\text{SO}}$ is the spin-orbit coupling Hamiltonian.
We notice that all wave vectors are replaced by $\bm{k}\to \bm{k}-e\bm{A}/$, where $\bm{A}=\pqty{-B_0 y, B_0 x, 0}/2$ is the vector potential.
As a bulk inversion-symmetric material, Silicon will only have Rashba spin-orbit coupling due to the presence of the gate electric field, which breaks the structural inversion symmetry \cite{Winkler2003}.
The leading order of the Rashba spin-orbit coupling term read:
\begin{equation}
    H_{\text{SO}} = \alpha_{\text{R}} (k_y \sigma_x - k_x \sigma_y) \,.
\end{equation}
$\alpha_{\text{R}}$ is the Rashba spin-orbit coupling coefficient, which can be derived from a Löwdin partitioning calculation of the standard $\bm{k}\cdot\bm{p}$ multi-band Kane model.
We neglect interface-induced (Dresselhaus-like) SOC due to amorphous Si/SiO${}_2$ interface, which therefore requires an individual device modeling.
With the application of an external magnetic field $\bm{B}_0$, the Zeeman Hamiltonian takes the form $H_{\text{Z}} = g \mu_B \bm{B}_0\cdot\bm{\sigma}/2$, where the electron Landé $g$-factor is 2, $\mu_{\text{B}}$ is the Bohr magneton, and $\bm{\sigma}=(\sigma_x, \sigma_y, \sigma_z)$ is the standard electron spin operator, with $\sigma_x, \sigma_y, \sigma_z$ being the Pauli matrices.
In this work, we assume the field is aligned along $\hat z$; thus, the Zeeman Hamiltonian reduces to $H_{\text{Z}} = g \mu_B B_z \sigma_z/2$.
Now, if we perform the diagonalization of Eq.~\eqref{Eq: QD total Hamiltonian}, we obtain the effective qubit Hamiltonian, which can be described by an effective two-level system $H_{\text{Qubit}}= \omega_0 \sigma_z/2$, where $\omega_0$ is the qubit Larmor frequency.

Next, we discuss the axion field and its coupling mechanism and its coupling to an electron spin qubit following Ref.~\cite{Alexander2018, Yi2023}.
The axion-electron interaction is considered under the assumption that the electron is in the low-energy regime under the non-relativistic approximation.
A free pseudo-scalar axion field $a(x)$ is governed by the Lagrangian density $\mathcal{L}_{a} = \partial^\mu a \partial_\mu a/2 - m_a^2 a^2/2$, where $m_a$ is the axion mass, and $a$ is the axion field.
Recent laboratory and astrophysical searches have set upper limits on QCD axions, excluding masses above $10^{-3}$~eV for DFSZ-type axion–electron coupling, while no firm lower bound exists and ALPs could extend down to $m_a \approx 10^{-22}$~eV \cite{Hu2021}.
In this work we focus on the axion window with mass from $10^{-6}$~eV to $10^{-3}$~eV, where resonant cavity and spin-based techniques offer competitive sensitivity \cite{Wang2022, Yang2024}.
The axion-electron interaction term is given by:
\begin{equation}
  \label{Eq: EA Interaction} 
  \mathcal{L} = \mathcal{L}_{e} + \mathcal{L}_{a} - i g_{ae} a(x) \bar{\psi}(x) \gamma^5 \psi(x) \,,
\end{equation}
where $\mathcal{L}_{e}=\bar{\psi}(x)\left(i \gamma^\mu \partial_\mu - m_e\right) \psi(x)$ is the Dirac electron Lagrangian density.
$\gamma^\mu$ are standard Dirac gamma matrices, $\psi$ and $\bar{\psi}$ are the electron Dirac field and its conjugate field, respectively.
To derive the effective axion-electron coupling Hamiltonian, we firstly transform Eq.~\eqref{Eq: EA Interaction} into corresponding Hamiltonian:
\begin{small}
\begin{equation}
    \label{Eq: AE Hamiltonian}
    H=\int \psi^{\dagger}(x)\left[-i \bm{\alpha} \cdot \nabla+\beta m_e + g_{ae} a(x) \gamma^5\right] \psi(x) \dd[3]x \,,
\end{equation}
\end{small}
where $\bm{\alpha}=\gamma_0 \bm{\gamma}, \beta=\gamma_0$.
Additionally, we denote $\beta m_e + g_{ae} a(x) \gamma^5$ as $W$.
Then we can perform a Schrieffer-Wolff transformation on Eq.~\eqref{Eq: AE Hamiltonian} up to second order of $W$ to obtain the effective Hamiltonian for the positive energy band, where $W_1=\beta m_e$, and $W_2=i g_{ae} a(x) \gamma^5$ respectively.
By constructing the operator $S=-i \beta W/(2m_e)$, we notice that $\{\beta, W\}=0$ and $\{\beta,S\}=0$.
Therefore, the unitary operator in the Schrieffer-Wolff transformation can be selected as $U_{\text{SW}}=e^{-iS}$.
Then our effective Hamiltonian $UHU^\dagger$ will read $H_{\text{ae}}=W^2/(2m_e)$, where the $W^2$ term can be expanded as:
\begin{equation}
    W^2 = W_1^2 + W_2^2 + \{W_1, W_2\} \,.
\end{equation}
Using the condition of slow varying axion field over the size of qubit (neglecting the diagonal energy shift), we can get:
\begin{equation}
    H_{\text{eff}} = \frac{i g_{ae}}{2m_e} \bqty{a(x)\bm{\sigma}\cdot\bm{p}+\bm{\sigma}\cdot\bm{p}a(x)}\,.
\end{equation}
By projecting this Hamiltonian on the qubit space, we have:
\begin{equation}
    H_{\text{int}} = \frac{g_{ae}}{2m_e} \bm{\sigma}\cdot\nabla a(x) \,.
\end{equation}
Now, the resulting qubit dynamics are described by the Hamiltonian:
\begin{equation}
    H_{\text{total}} =  \omega_0 \sigma_z + \frac{g_{ae}}{2m_e} \bm{\sigma}\cdot\nabla a(x) \,.
\end{equation}
In this effective Hamiltonian, $\omega_0$ incorporates all the corrections to the energy levels including the axion-electron interactions, and the spin-orbit coupling corrections.

The axion field can be treated as a classical oscillating field expressed as $a(t,x)=a_0 \cos(m_a t - \bm{k}\cdot\bm{x}+\varphi)$, where $\varphi$ is an arbitrary phase factor.
We apply a static magnetic field $B_0 \hat{\mathbf{z}}$ to set the qubit spin-quantization axis along the device z-direction.
We notice that, in general, the axion-induced effective magnetic field $\bm{B}_{\text{eff}}$ does not have a preferred orientation and may not align with the device z-direction.
However, only the projection of $\bm{B}_{\text{eff}}$ contributes to the Zeeman splitting $g \mu_B B_{\text{eff}}$ and thus accumulates a detectable phase shift in Ramsey-type measurements. 
All transverse components merely drive spin flips and do not produce phase shift in the $\sigma_z$ basis.
To maximize sensitivity, one can deploy multiple qubit arrays along different axes or mount the device on a rotatable stage, thereby tracking the expected axion-wind direction and optimizing the projection factor $\cos\theta_0$.
In the following discussions, we restrict our analysis to the z-direction component.

The total Hamiltonian takes the form:
\begin{equation}
    \label{Eq: total_H}
    H_{\text{total}} = \frac{\omega_0}{2} \sigma_z + \frac{g_{ae}}{2m_e} k_z a_0 \sin(m_a t-k_z z + \varphi) \sigma_z \,,
\end{equation}
The axion momentum is defined as $k_z = m_a v$, where $v \approx 10^{-3}c$ is the typical virial velocity of halo dark matter.
The amplitude of the axion field $a_0$ is related to the local dark matter energy density $\rho_a$ via $a_0= \sqrt{2\rho_a}/m_a$.
Comparing with Eq.~\eqref{Eq: total_H} and Zeeman term, we define:
\begin{equation}
  \label{Eq: axion magnetic field}
  B_{\text{eff}} = \frac{g_{ae}}{\gamma_e} v \sqrt{2 \rho_a} \,.
\end{equation}
To analyze the spin polarization dynamics, we construct the time-evolution operator, denoted by $U=\operatorname{diag}[e^{-i\phi_+t}, e^{-i\phi_-t}]$, where:
\begin{small}
\begin{equation}
  \phi_{+}(t)=\int_0^t E_{+}\left(t^{\prime}\right) \dd t^{\prime} \quad \phi_{-}(t)
  =\int_0^t E_{-}\left(t^{\prime}\right) \dd t^{\prime} \,.
\end{equation}
\end{small}
The energy eigenvalues for the two-level system are:
\begin{equation}
  E_\pm = \pm \bqty{\frac{\omega_0}{2} + \frac{g_{ae}v}{2m_e} \sqrt{2\rho_a} \sin(m_a t-k_z z + \varphi)} \,.
\end{equation}
At $t=0$, we prepare $\ket{\psi(0)} = c_+(0) \ket{\uparrow} + c_-(0)\ket{\downarrow}$, with $c_+(0)$ and $c_-(0)$ are arbitrary.
Evolving under $E_\pm$, the state can be expressed as $\ket{\psi(t)}=c_+(0)e^{-i\phi_+(t)}\ket{\uparrow}+c_-(0)e^{-i\phi_-(t)}\ket{\downarrow}$.
Now we can evaluate the spin polarization expectations as a function of $t$, for the $\sigma_z$ component $\expval{\sigma_z(t)}=\mel**{\psi(t)}{\sigma_z}{\psi(t)}$, which is $\expval{\sigma_z(t)}=\norm{c_+}^2-\norm{c_-}^2$, a constant.
For $\expval{\sigma_x(t)}$ term, we have $\expval{\sigma_x(t)}= 2 \Re\bqty{c_+^* c_- e^{-i\Delta \phi(t)}}$, where $\Delta \phi(t)$:
\begin{equation}
    -\omega_0 t - \frac{g_{ae}v\sqrt{2\rho_a}}{2m_e\omega_a} (\cos(m_a t-k_z z + \varphi)-\cos\varphi) \,.
\end{equation}   
If we consider the initialization of the state is $\ket{\psi(0)}=(\ket{\uparrow}+\ket{\downarrow})/\sqrt{2}$, the $\expval{\sigma_z(t)}$ term will be $0$.
The $\expval{\sigma_x(t)}$ read:
\begin{small}
\begin{equation}
  \label{Eq: Exp sigma x}
  \cos(\omega_0 t + \frac{g_{ae}v\sqrt{2\rho_a}}{2m_e\omega_a} (\cos(m_a t-k_z z + \varphi)-\cos\varphi))
\end{equation}    
\end{small}
In this simple calculations, the absence of off-diagonal deriving terms prevents conventional spin Rabi oscillations as seen in the electron spin resonance.
Instead, the accumulated phase of the spin polarization directly tracks the axion field oscillation frequency.
Eq.~\eqref{Eq: Exp sigma x} indicates that, in addition to the standard Larmor precession at frequency $\omega_0$ along the z-direction, the axion field will induce a small sinusoidal modulation.
This sets the foundation of searching the axion field signals using electron spin qubits.

\subsection{SNR estimation}
\begin{figure}[htb!]
    \centering
    \includegraphics[width=3.14in, keepaspectratio]{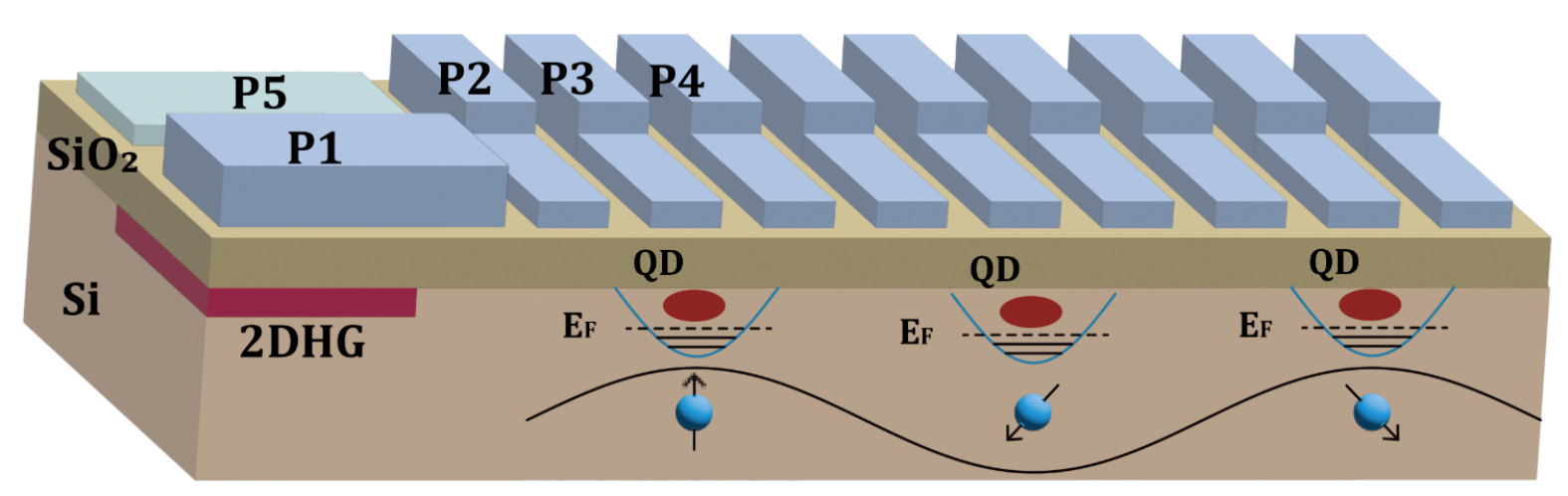}
    \caption{
        Schematic planar silicon quantum dot.
        In this specific design, we focus on a quantum dot array in the silicon layer.
        By applying a gate electric field $F_z$ via gate P1, holes accumulate in silicon and are confined vertically against the silicon oxide (indicated at the location of the two-dimensional hole gas).
        The single quantum dot is formed using gates P2--P5.
        The gates P2 and P4 provide confinement in the $\hat{x}$-direction, while gate P5 provides confinement in the $y$-direction. 
        P3 is used as the top gate of the quantum dot, accumulating a single hole in the potential well beneath.
        The resulting potential is indicated schematically below the gates.
        Then the quantum dot array repeats the set-up for P2--P4.
    }
    \label{fig:QuantumArray}
\end{figure}
We now consider an ensemble of $N$ independent electron spins qubit subjected to the oscillating axion field.
Over each spin coherence time $T_2$, it accumulates a phase shift $\phi_{\text{sig}} = \gamma_e B_{\text{eff}} T_2$.
Recent advances on application-specific integrated circuits integrated analogue-to-digital conversion (ASIC ADC), particularly the development of Skipper-type architectures, have enabled single-electron sensitivity with effectively zero readout noise \cite{Giardino2022, Kiene2025}.
By performing multiple non-destructive charge measurements on the same sensing element, Skipper ADCs can suppress the readout noise to sub-electron levels to extract the spin polarizations by spin–to–charge conversion.
As a result, in such measurement platforms, though the magnetic field $\delta B=\eta_B/\sqrt{T_2}$ for a single spin contributes an uncertainty
\begin{equation}
    \Delta\phi = \gamma_e \delta B T_2=\gamma_e \eta_B\sqrt{T_2} \,.
\end{equation}
After repeating $M$ the measurement over a total integration time $t$ improves the SNR by $M=t/T_\text{2}$, leading to the array total SNR:
\begin{equation}
    \text{SNR} = \frac{\phi_{\text{sig}}}{\Delta\phi}\sqrt{M} =\frac{\gamma_e B_{\text{eff}}T_2}{\gamma_e \eta_B \sqrt{T_2}}\sqrt{\frac{t}{T_2}}=\frac{B_{\text{eff}}\sqrt{t}}{\eta_B} \,.
\end{equation}
Embedding the quantum dot arrays in a high-$Q$ resonator tuned to the axion frequency $\omega_a \approx m_a / 2\pi$, the signal field amplitude is coherently enhanced by a factor of $Q$, assuming the signal lies within the cavity bandwidth. Thus, the enhanced SNR will multiply with a factor $Q$, using Eq.~\eqref{Eq: axion magnetic field}, the SNR in terms of axion parameters is $B_\text{eff}=g_{ae} v \sqrt{2 \rho_a}/\gamma_e$.
This gives the standard expression of SNR in decibels $\text{SNR}_{\text{dB}}=20 \log_{10}(\text{SNR}_{\text{amp}})$.
This expression clearly shows that detection sensitivity improves with larger spin number $N$, longer coherence and integration times, higher resonator quality factor $Q$.
Further quantum enhancement (e.g., via full-entanglement) can replace $\sqrt{N}$ by $N$, yielding the Heisenberg limit as
\begin{equation}
    \label{Eq: SNR_en}
    \text{SNR}_{\text{entangled}} = \frac{Q g_{ae} v \sqrt{2\rho_a} N\sqrt{t}}{\gamma_e\eta_B} \,.
\end{equation}

\subsection{Noise simulation and signal processing protocol}
\label{S2: SS3 Noise simulation and signal processing protocol}
To maintain a clear signal of the axion field, encoded in the envelope of the qubit Larmor frequency, we need to consider the mitigation of the electrical noise at the device design level.
In silicon quantum-dot spin qubits, the dominant electrical noise arises from single-charge defects at the Si/SiO${}_2$ interface.
Dipolar two-level defects also exist but contribute only marginally and are therefore neglected here  \cite{Culcer2013}.
A single charge defect can be modeled by a screened Coulomb potential:
\begin{equation}
    V_{\text{scr}}(\bm{x})=\frac{e^2}{2 \epsilon_0 \epsilon_r (2\pi)^2} \int_0^{2 k_F} \frac{e^{-i \boldsymbol{q} \cdot \bm{x}}}{q+q_{T F}} \dd[2]{q}\,.
\end{equation}
$\epsilon_0$ is the vacuum permittivity, $\epsilon_r=11.68$ is the relative permittivity of silicon.
The Thomas-Fermi wave-vector is $q_{\text{TF}}=0.97$~nm${}^{-1}$, and $k_F$ is the Fermi wave-vector selected to be 0.1~nm${}^{-1}$, according to Ref.~\cite{Bermeister2014}.
In our case, we are concerned with the noise power spectrum due to the ensemble of the charge defects \cite{Uren1985}.
Considering a single random telegraph noise source, the power spectral density is:
\begin{equation}
    S_\delta^{\text{RTN}}(\omega)=\frac{\delta \varepsilon_Z^2 \tau}{1+\omega^2 \tau^2} \,.
\end{equation}
$\delta \varepsilon_{\text{Z}}^2$ quantifies the energy correction to the qubit energy splitting due to a single charge defect.
To evaluate this term, we first diagonalize the Hamiltonian $H_{\text{total}} + V_{\text{scr}}$ to obtain $\omega_1$, the qubit Larmor frequency in the presence of a charge defect.
We then define $\delta \varepsilon_{\text{Z}} = \omega_1 - \omega_0$, where $\omega_0$ is the qubit Larmor frequency without the defect, as calculated in Eq.~\eqref{Eq: QD total Hamiltonian}.
$\tau$ is the switching time of the random telegraph noise source.
The range of $\tau$ (from $\tau_{\text{min}}$ to $\tau_{\text{max}}$) can be determined experimentally, as in Ref.~\cite{Connors2019}.
Following the derivation of Ref.~\cite{Wang2025}, we average $\tau$ to obtain a $1/f$ noise power spectral density of the form $S(\omega)=\alpha \expval{\delta \varepsilon_{\text{Z}}^2}/\omega$, where $\alpha=0.8$ for our model.
This $1/f$ noise spectrum forms the basis for subsequent filtering steps, which we discuss in the following sections.

To make the best use of the potential of quantum sensing in quantum dot array platform, we discuss a possible blueprint for the experimental setup to extend the lower boundary of the qubit-axion couplings and the effective magnetic field.
A starting point is the dispersive coupling (with $g_{j,\text{bare}} \ll|\Delta_j|$, $\Delta_j=\omega_{01}^{(j)}-\omega_c$, and $\omega_{01}^{(j)}$ the Larmor frequency of the $j$-th qubit) between $N$ semiconductor spin qubits and a single mode of a high-$Q$ microwave resonator at frequency $\omega_c$~\cite{Kuruma2020,Bottcher2022}.  
Although current CMOS technology is still maturing toward large-scale entanglement and should not be over-interpreted, recent proposals have demonstrated industrial‐scale quantum‐dot arrays~\cite{Elsayed2024,Kiene2025}.

The simplified multi‐qubit–resonator Hamiltonian in the dispersive regime is
\begin{equation}
    H 
    = \sum_{j=1}^N \frac{\omega_{01}^{(j)}}{2}\,\sigma_z^{(j)}
      + \omega_c\,a^{\dagger}a
      + \sum_{j=1}^N \chi_j\,\sigma_z^{(j)}\,a^{\dagger}a
    \,,
\end{equation}
where $a$ ($a^\dagger$) is the cavity‐mode annihilation (creation) operator, and $ \chi_j \equiv g_{j,\text{bare}}^2 / \Delta_j$.

A global microwave drive at frequency $\omega_d\approx\omega_{01}$, 
injected through the cavity input port, prepares the qubit array in the state
\begin{equation}
    \ket{\psi_0}
    = \bigotimes_{j=1}^N \frac{\ket{0}_j + \ket{1}_j}{\sqrt{2}} \,
\end{equation}
An external magnetic field $B_0$ along the $z$-axis sets the qubit Larmor 
frequencies $\{\omega_{01}^{(j)}\}$. We then measure the collective spin 
polarizations
\begin{equation}
    \langle S_x\rangle = \sum_{j=1}^N \langle\sigma_x^{(j)}\rangle
    \,,\quad
    \langle S_z\rangle = \sum_{j=1}^N \langle\sigma_z^{(j)}\rangle \,,
\end{equation}
whose time-dependent modulations encode the axion–qubit interaction.
After the interaction period, the RF reflectometry signal from the gate‐based LC resonator on the CMOS quantum dot array is routed into a cryogenic CMOS readout ASIC, which co‐packages:
1. a two‐stage SiGe/CMOS LNA operating at the 4K stage with a noise temperature $T_N\lesssim4\,$K (with future goals of $T_N\lesssim1\,$K).
2. a cryo‐CMOS IQ down‐converter (mixer + on‐chip narrow-band IF filters) referenced to the qubit Larmor frequency $\omega_{01}$.
3. and an in‐pixel Skipper‐inspired ADC (65nm CMOS) that performs $M$ non‐destructive charge‐sampling (``pile‐up") in the analogue domain before a 10‐bit SAR conversion at  about $100$~kS/s, achieving sub–single‐electron noise ($0.48e^-$rms) at $T\approx100$~K for spin–to–charge conversion.
The serialized digital output of this ASIC is sent over a low‐power link to a room‐temperature DAQ system, where the further process is applied.
one branch reconstructs and FFT‐processes the demodulated time‐domain waveform for broadband spectral analysis, and the other ingests the multi‐sampled ADC stream for high‐resolution, narrowband detection of spin‐resonance shifts.
By integrating the LNA, down‐converter and multi‐sampling ADC on a single cryo‐CMOS chip directly adjacent to the quantum dot sensor (rather than placing the ADC at room temperature), parasitic loss and thermal noise are minimized while preserving sub–single‐electron resolution.
\begin{figure}[t!]
    \centering
    \includegraphics[width=3.14in, keepaspectratio]{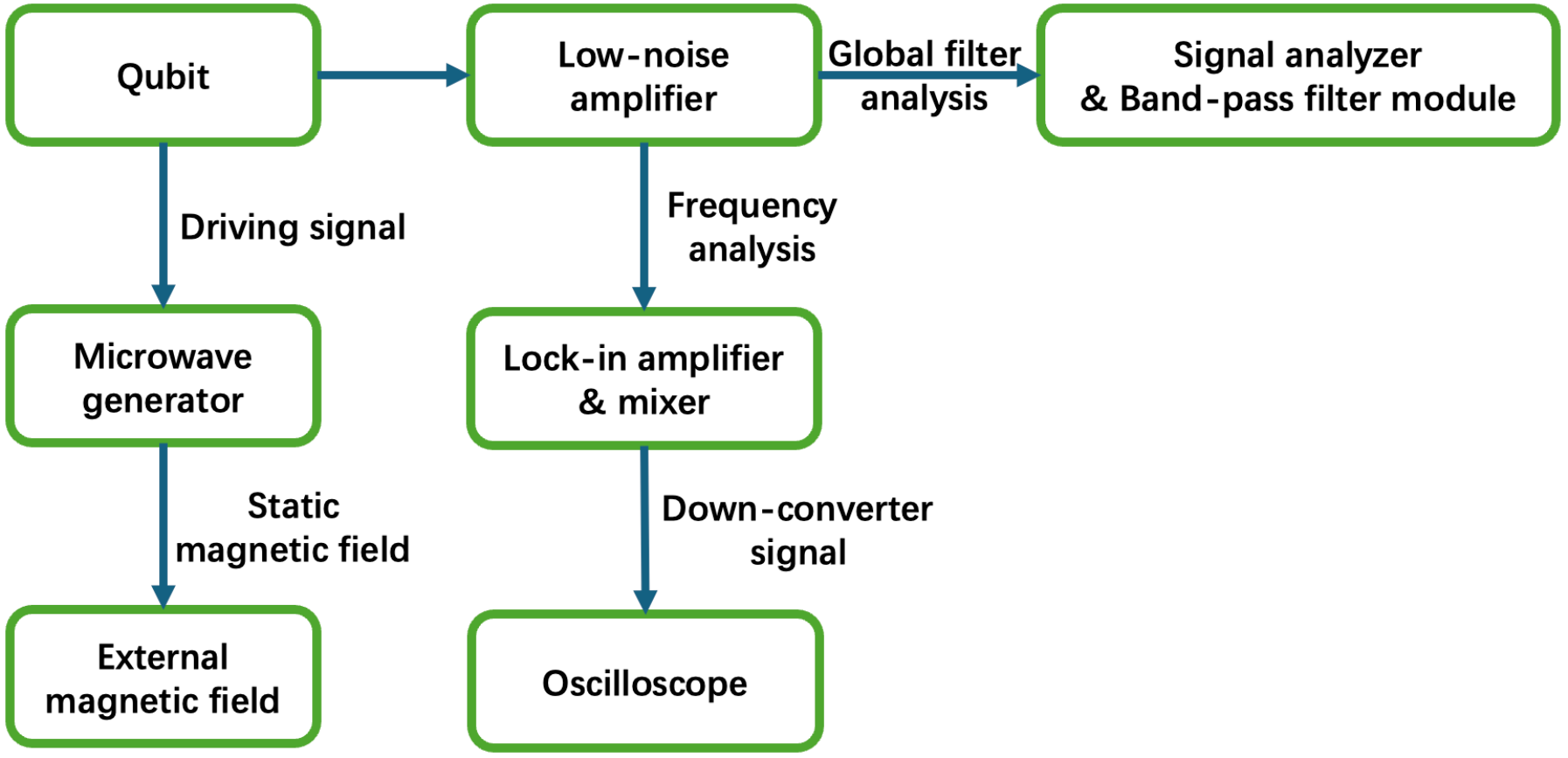}
    \caption{
        A possible workflow combining FFT analysis and lock-in detection.
        The LNA outputs signals to both the FFT module for global frequency analysis 
        and the Lock-in Amplifier for specific frequency detection.
    }
    \label{fig:setup}
\end{figure}

\section{Results and discussion}
\label{S3: Results and discussion}
\begin{figure}[htbp!]
    \centering
    \includegraphics[width=3.14in, keepaspectratio]{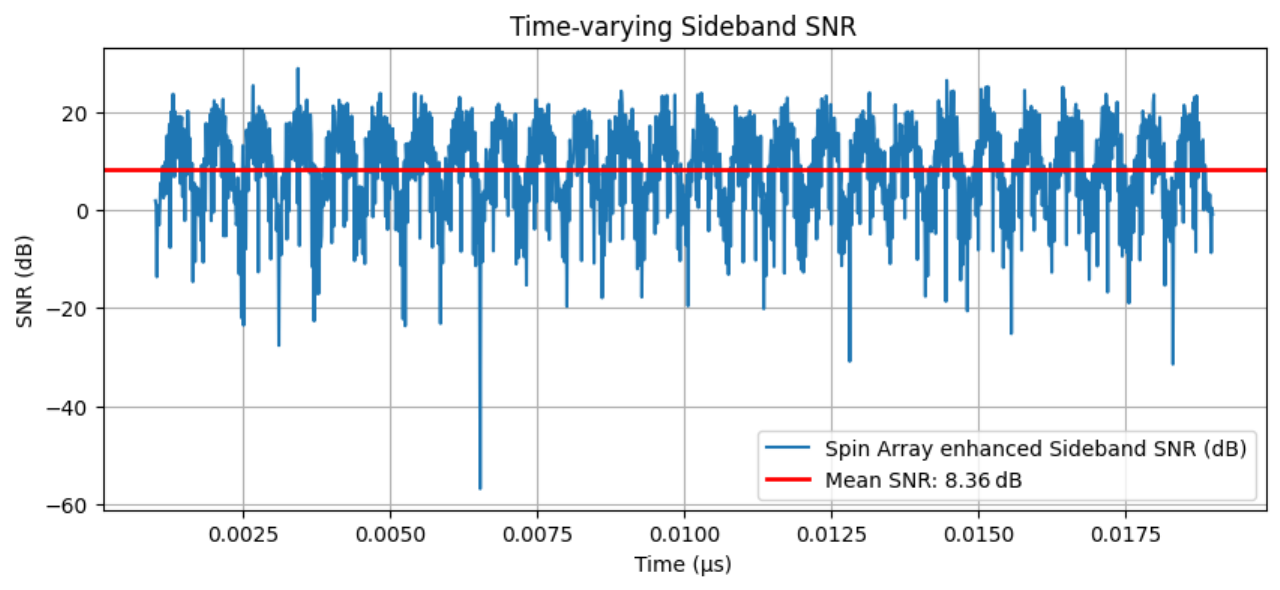}
    \caption{
        Time-dependent SNR of the filtered transverse spin response $\langle\sigma_x\rangle$,
        computed over sliding windows of size 100.
        The SNR fluctuates between about $-60$ dB and 20 dB due to oscillations, spectral leakage, and drift,
        while remaining above 8.36 dB, demonstrating robust detectability of axion-induced modulation.
        The best-case simulated SNR reaches 22.5 dB according to
        $\text{SNR}=20\log_{10}(Q N \sqrt{t} B_{\text{eff}} / \eta_B)$.
    }
    \label{fig:snr}
\end{figure}

In this section, we firstly discuss the possibility of scanning axion induced magnetic field based on our quantum dot array and resonator architectures.
In our simulation, for computational efficiency, the target axion mass is of order $\times 10^{-6}$~eV, our simulations yield predicted sidebands at $\pm$720~MHz corresponding to an axion mass of 3~$\mu$eV.
Recall the qubit axion interaction Hamiltonian derived in Eq.~\eqref{Eq: total_H}, the axion induced magnetic field enters the qubit Hamiltonian as a time‐dependent detuning,  
\begin{equation}
    H(t) = \frac{\hbar}{2}\bigl[\omega_0 + \delta\omega\cos(\omega_a t)\bigr]\sigma_z + \cdots \,,
\end{equation} 
where $\delta\omega$ is the modulation amplitude and $\omega_a$ the axion frequency.
In the frequency domain, such a frequency modulation does not deposit energy at the carrier $\omega_0$ itself but rather redistributes it into sidebands at $\omega_0 \pm n\,\omega_a$, where $n=1,2,\cdots$.  
This relative amplitudes are given by Bessel functions $J_n(\beta)$, where $\beta=\delta\omega/\omega_a$ is the modulation index.
The oscillatory detuning mixes the qubit transitions with the axion oscillations, generating spectral lines at the sum and difference frequencies.
Consequently, this coupling results in the appearance of sidebands at frequencies $f_{\text{main}} \pm n f_{\text{axion}}$, where $f_{\text{main}}$ is the main qubit frequency, $f_{\text{axion}}$ is the axion frequency.
Then the sideband is given by:
\begin{equation}
    f_{\text{sideband}} = f_{\text{main}} \pm n f_{\text{axion}}, \quad n \in \mathbb{Z}
    \label{Eq: modulation}
\end{equation}
The sideband modulation index $\beta$ is defined as the ratio of the peak frequency deviation to the modulation frequency:
\begin{equation}
  \label{Eq: beta_def}
  \beta = \frac{\Delta\omega_{\text{max}}}{\omega_a}=\frac{\gamma_eB_\text{eff}}{\omega_a} \,.
\end{equation}
Using Eq.~\eqref{Eq: axion magnetic field}, we can evaluate the $\beta$ as $\beta=g_{ae} v \sqrt{2\rho_a}/m_a$.
Using the typical axion parameters, we use $m_a=10^{-6}$~eV and $\rho_a=0.3$~GeV/cm${}^3$ and $g_{ae}=10^{-13}$, the modulated index can be estimated as $\beta \approx 10^{-21}$.
We notice that this value is still marginally small, however, once we taking the advantage of the qubit array with a magnetic sensitivity $\eta_B \approx 100$ nT/$\sqrt{\text{Hz}}$ \cite{Szechenyi2017}, the lower boundary $\beta_{\text{min}}$ can cover this region. The expression  $\beta_{\text{min}}$ reads $\beta_{\text{min}} = \gamma_e \eta_B/(\omega_a \sqrt{T}_2)$.
After high-$Q$ resonant and entanglement enhancement, $\beta_{\text{min}}$ can be improved to:
\begin{equation}
    \beta_{\min}(\text{Enhanced)}=\beta_{\text{min}}\frac{1}{QN\sqrt{M}}=\frac{\gamma_e \eta_ B}{QN\sqrt{t}\omega_a} \,,
\end{equation}
where we can see that at $Q=1$, $\beta_{\text{min}}\approx10^{-17}$ and at $Q=10^5$, $\beta_{\text{min}}\approx10^{-22}$, where $Q=10^5$ are selected from Ref.~\cite{Holman2020}. For a valuable result, $5\sigma$ is required for the detection as $\text{SNR}_{\text{amp,th}}=\sqrt{25\sigma_{n}^2/\sigma_{n}^2}=5$.
With the enhancement of the experimental setup, $\text{SNR}_{\text{amp, th}} = Q N\sqrt{t}B_{\text{eff}}/\eta_B$, under the best condition, it can reach 22.5 as seen Fig.~\ref{fig:snr}.
This result agrees with our previous attempt via phase uncertainty in Eq.\eqref{Eq: SNR_en} and gives the same form, then the searching sensitivity must follow the threshold below:
\begin{equation}
   \beta(g_{ae}, \omega_a)\geq5 \beta_{\text{min}} \,.
\end{equation}
\begin{figure}[htbp!]
    \centering
    \includegraphics[width=3.14in, keepaspectratio]{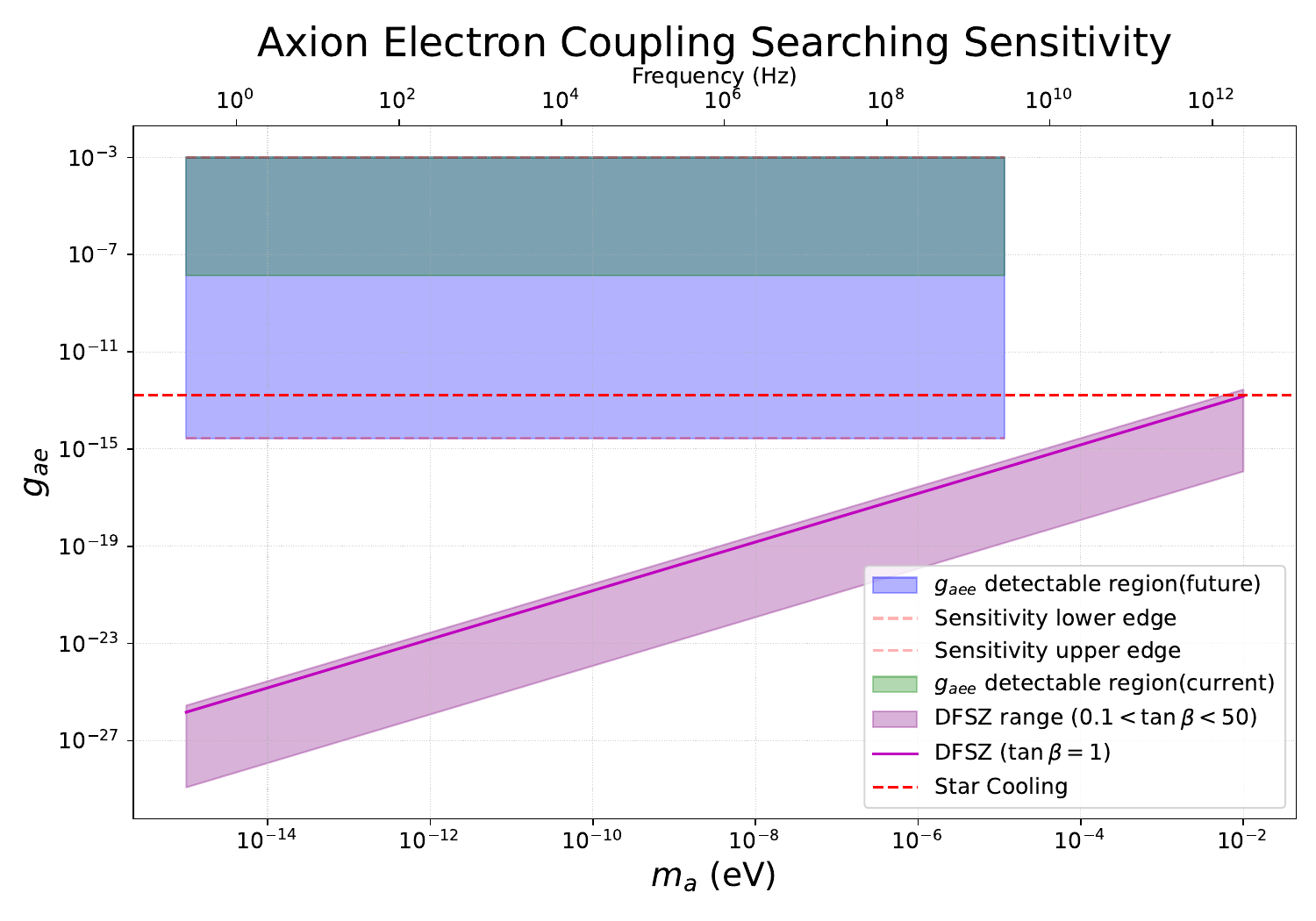}
    \caption{
        Projected search sensitivity for the axion--electron coupling constant $g_{ae}$ 
        as a function of axion mass $m_a$ (bottom axis) and corresponding frequency (top axis).
        The shaded purple region indicates the detectable parameter space with next-generation devices 
        ($Q=10^{6}$, $N=10^{6}$, $T_{2}=100$ ms), bounded by the dashed red line. 
        The green region shows the space of current devices ($Q=10^{4}$, $N=16$, $T_{2}=1$ ms). 
        The purple band denotes the DFSZ model range ($0.1<\tan\beta_a<50$), with the solid purple line 
        corresponding to $\tan\beta_a=1$. 
        The red dotted line indicates current stellar cooling limits \cite{Oda2020}.
    }
    \label{fig:Sensitivity}
\end{figure}

Although our present sensitivity (Fig.~\ref{fig:Sensitivity}) does not yet fully cover the classical DFSZ band for QCD axions, the consideration of CMOS quantum‐dot spin platform remains highly significant for several reasons.
First, any ALPs with an electron coupling $g_{ae}$ above our sensitivity threshold can be probed, including models with suppressed photon couplings or mass scales outside the QCD axion window.
Second, the cryo-CMOS readout ASIC, co-packaged with the quantum-dot array, can be tiled to $10^4$–$10^6$ pixels on a single 300~mm wafer, enabling future detectors with $N \approx 10^8$–$10^{10}$ spins.
Here we also sketch a possible roadmap to further improve the performance.
1. Advances in materials and spin control are pushing quantum-dot $T_2$ from the microsecond to the millisecond scale.
2. New high-$Q$ three-dimensional and photonic-crystal resonator designs promise field-enhancement factors $Q \gtrsim 10^6$.
3. Next-generation amplifier like JPAs/TWPAs and Skipper-inspired CMOS ADCs are on track to achieve noise temperatures $T_N < 0.1$~K and sub–single-electron resolution at about 100~kS/s.
With these developments, we anticipate that a future device incorporating $N\gtrsim10^8$ spins, millisecond coherence, and megahertz-bandwidth high-$Q$ readout will extend sensitivity well into—and ultimately beyond the DFSZ parameter region.

\begin{figure}[htbp!]
    \centering
    \includegraphics[width=3.14in, keepaspectratio]{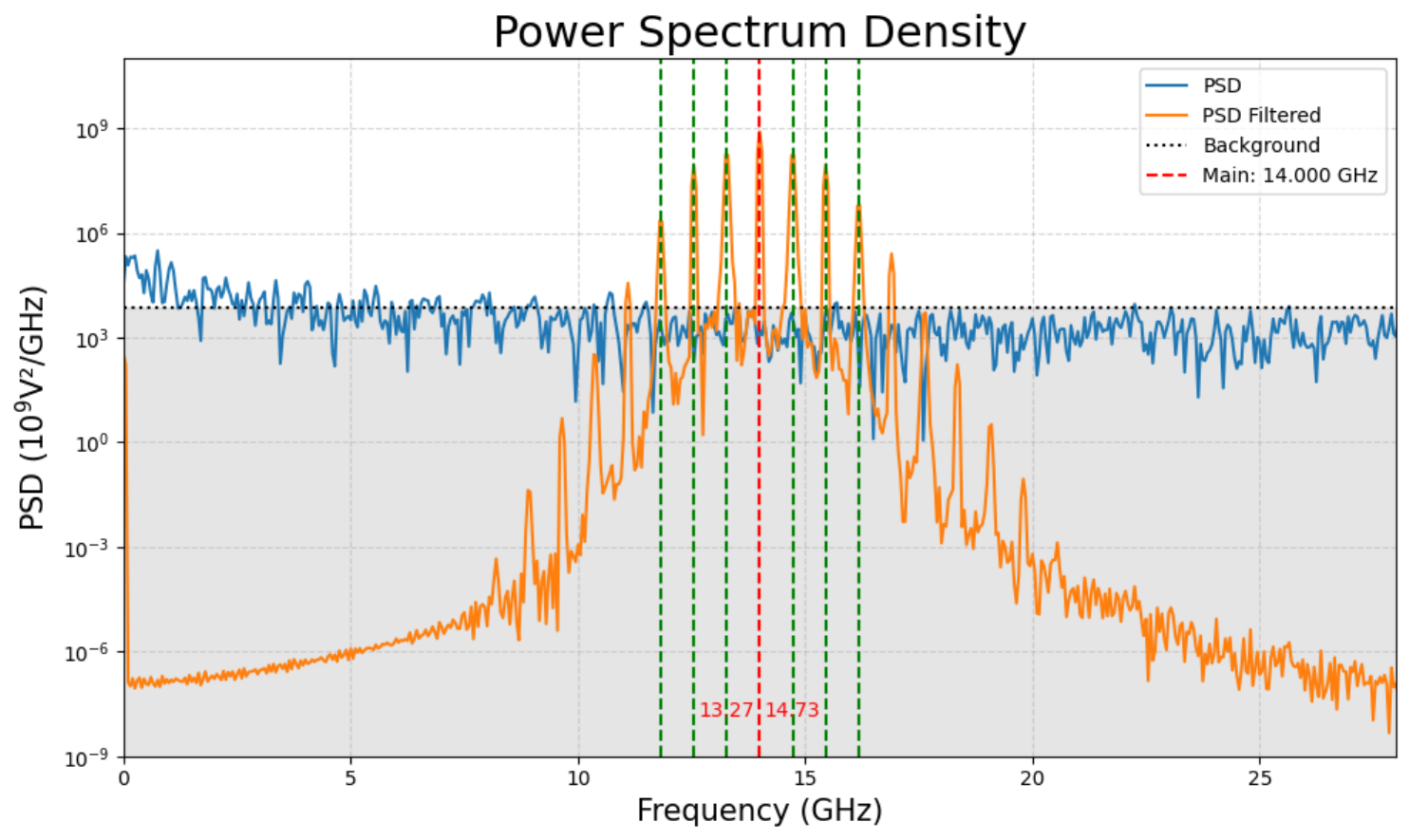}
    \caption{
        Power spectral density (PSD) of the received signal with an input SNR of $2.73$ dB before filtering.
        A prominent carrier peak appears at 14 GHz, with symmetric sidebands at 13.27 and 14.73 GHz, consistent with axion-induced modulation.
        The noise floor is shown as a reference for signal integrity and spectral purity.
    }
    \label{fig:Power_spectrum}
\end{figure}

For our reference numerical simulation, the experimental parameters are set up as in the Tab.~\ref{tab:sim-parameters} for better visualization, the main frequency $f_{\text{main}}$ for the electron-spin qubit under $B_0=0.5$~T is $14$~GHz, while the axion frequency $f_{\text{axion}}$ is approximately $720$~MHz, and the amplitude has been scaled by the factor in Eq.~\ref{Eq: SNR_en}.
Consequently, sidebands are expected at $f_{\text{sideband,1}} = 14\pm 0.72$~GHz, $f_{\text{sideband,2}} = 14 \pm 2 \times 0.72$~GHz, and so on, corresponding to the first, second, and higher-order sidebands.
In the presence of noise, these sidebands are often obscured by undesired signal components outside the frequency range of interest.
To isolate the axion-induced signal, we employ a band-pass filter with a pass-band designed around $f_{\text{main}}\pm2\times f_{\text{axion}}$.
This filter is designed with the following cut-off frequencies:
\begin{equation}
  f_{\text{low}} = 0.9 \times f_{\text{main}}, \quad f_{\text{high}} = 1.1 \times f_{\text{main}} \,.
\end{equation}
\begin{table}[htbp]
  \caption{\label{tab:sim-parameters}Simulation parameters used in this work.}
  \begin{ruledtabular}
  \begin{tabular}{lll}
    Symbol & Description & Value \\
    \hline
    $m_a$           & Axion mass                       & $3\times10^{-6}$ eV \\
    $\rho_a$        & Local dark-matter density        & 0.3 GeV cm$^{-3}$ \\
    $v_a$           & Axion velocity                   & $1\times10^{-3}\,c$ \\
    $\gamma$        & Spin gyromagnetic ratio          & $28\times10^{9}$ Hz/T \\
    $T_1$           & Longitudinal relaxation time     & 1 ms \\
    $T_2$           & Transverse dephasing time        & 100 $\mu$s \\
    $B_0$           & Static magnetic field            & 0.5 T \\
    $g_{ae}$        & Axion–electron coupling strength & $10^{-13}$ \\
    $Q$             & Quality factor (resonator)       & $10^{5}$ \\
    $N$             & Number of spins                  & $10^6$ \\
    $T_{\text{total}}$ & Total simulation duration     & 0.18 $\mu$s \\
    $\Delta t$      & Simulation time step             & $2\times10^{-3}$ ns \\
    Direction       & Axion direction                  & $z$-direction \\
  \end{tabular}
  \end{ruledtabular}
\end{table}
A realistic noise model (cf.\ Table~\ref{tab:sim-parameters}) is indispensable for our simulations because spin-qubit experiments inevitably experience a combination of broadband electronic noise, such as low‐frequency $1/f$ fluctuations, quantization errors, discrete two‐level system jumps, thermal drifts, and laboratory line‐frequency pickup \cite{Chan2018, Burkard2023}.
Accordingly, we set the white noise amplitude to $1\times10^{-3}$ and the pink noise amplitude to $2\times10^{-2}$ (with $f_{\min}=1$~Hz) to replicate measured spectral densities of room‐temperature amplifiers and device flicker noise; the readout noise parameters ($\sigma=2\times10^{-3}$, $p_{\text{spike}}=5\times10^{-3}$) to model digitization uncertainty and rare transient glitches in the data acquisition chain; the telegraph noise amplitude of $5\times10^{-4}$ and drift amplitude of $2\times10^{-4}$ to account for charge fluctuators and slow thermal variations in cryogenic components; and a modulated AC term ($A=1\times10^{-4}$ at $f_{\text{mod}}=50$~Hz) to include residual coupling to the power‐line environment.
And the Johnson–Nyquist noise has the power density of $4k_{B}TR$ with $T=300$~K.
These empirically motivated parameter choices ensure that our sensitivity estimates and SNR predictions accurately reflect the complex noise landscape encountered in high‐precision axion search experiments.
\begin{table}[htbp]
  \caption{\label{tab:noise-config}Simulation noise configuration parameters.}
  \begin{ruledtabular}
  \begin{tabular}{llllll}
    Noise type   & Amplitude        & $f_{\min}$ (Hz) & $p_{\text{spike}}$ & $f_{\text{mod}}$ (Hz) & $T$ (K) \\
    \hline
    White        & $1\times10^{-3}$ & –               & –                  & –                     & –     \\
    Pink         & $2\times10^{-2}$ & 1               & –                  & –                     & –     \\
    Readout      & $2\times10^{-3}$ & –               & $5\times10^{-3}$   & –                     & –     \\
    Telegraph    & $5\times10^{-4}$ & –               & –                  & –                     & –     \\
    Drift        & $2\times10^{-4}$ & –               & –                  & –                     & –     \\
    Modulated AC & $1\times10^{-4}$ & –               & –                  & 50                    & –     \\
    Thermal      & –                & –               & –                  & –                     & 300   \\
  \end{tabular}
  \end{ruledtabular}
\end{table}
For the power spectral density (PSD) plot, we expect to see several sidebands with different orders on the axion modulation frequency that follows our theoretical prediction in Eq.~\ref{Eq: modulation}.
Due to the small value of $B_{\text{eff}}$, we assume the Skipper ADC enhanced cryogenic experimental apparatus and high-$Q$ resonator coupled large $N$ qubit array could reduce the uncertainty together with time-integration $t$.
And our filter is designed to keep the first and second sideband for examination and noise reduction.
The filtered signal significantly reduces background noise and highlights the axion-induced sidebands, as illustrated in the PSD plots in Fig.~\ref{fig:Power_spectrum}.

Application of the band-pass filter can also separate the sideband signal from the noise floor, thereby enhancing the time-domain dynamic SNR but not the frequency near the sideband, so we have visualized the dynamic SNR around the sideband frequency at $13.27$~GHz for the qubit array, where the dynamic SNR can be evaluated as:
\begin{equation}
    \text{SNR}_{\text{dynamic}} = 10 \log_{10}\left(\frac{\operatorname{Var}(P_{\text{signal(sideband)}})}{\operatorname{Var}(P_{\text{noise(sideband)}})}\right),
\end{equation}
where $P_{\text{signal}} $ is the power at the signal frequency (e.g., near the sideband), and $P_{\text{noise}} $ is the power in the sideband region away from the background.
Our analysis of the dynamic output enhanced SNR over time reveals periodic fluctuations corresponding to the axion-induced modulations in Fig.~\ref{fig:snr}.

These fluctuations provide further evidence of axion influence on the qubit array, supporting the hypothesis that the axion field is coupled to the Hamiltonian of the electron spin qubits. 
\begin{figure}[htbp!]
    \centering
    \includegraphics[width=3.14in, keepaspectratio]{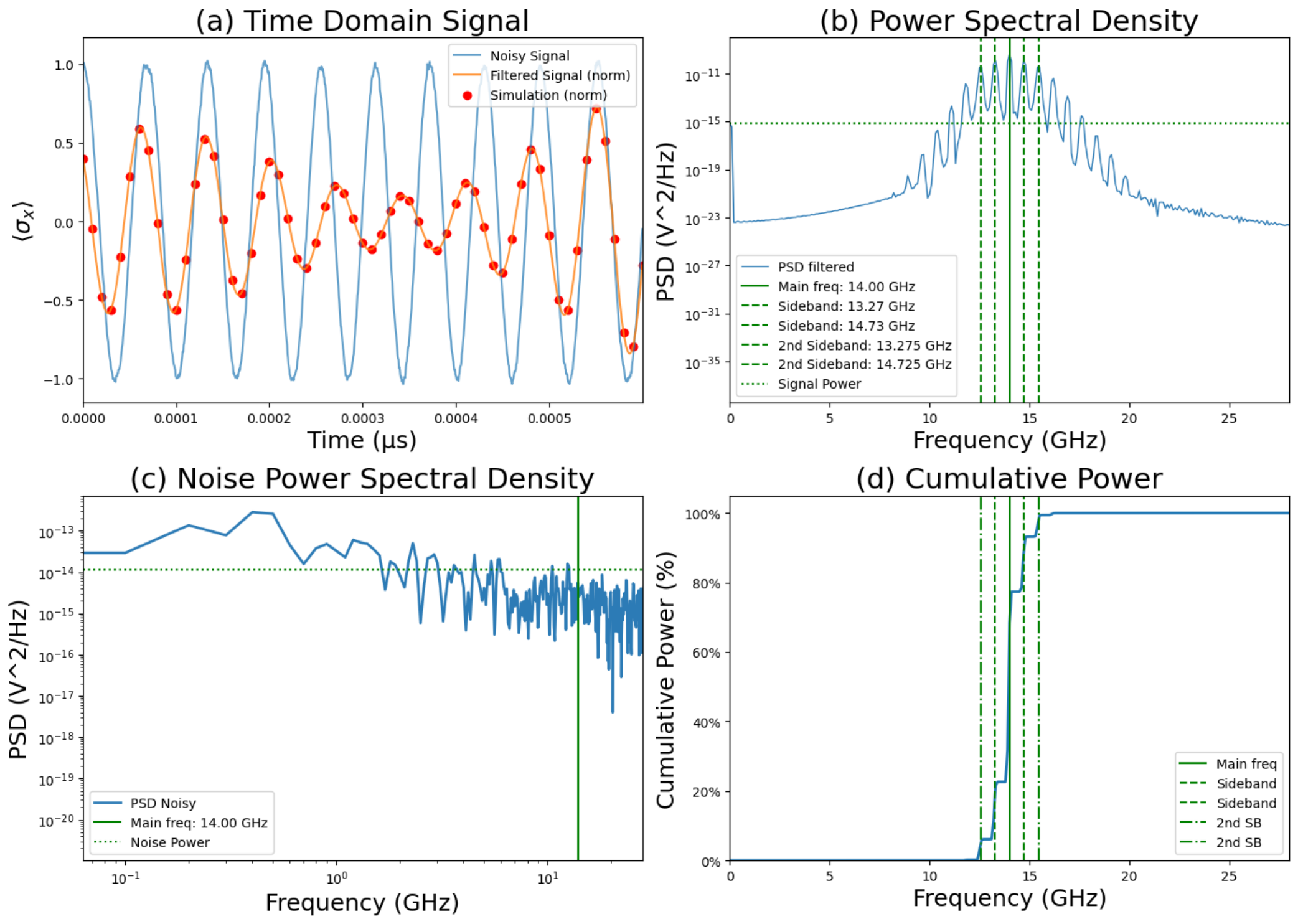}
    \caption{
        (a) Time-domain response of $\langle\sigma_x\rangle$ under axion coupling:
        raw noisy signal (blue), band-pass filtered output (orange), and discrete samples (red).
        (b) Power spectral density (PSD) of the filtered signal, showing the carrier at 14 GHz (solid green) 
        and sidebands at 13.27 and 14.73 GHz (dashed green).
        (c) Modeled noise PSD (blue) compared with the noise floor (horizontal dotted green) 
        and the carrier frequency (vertical green).
        (d) Normalized cumulative power versus frequency, with each step marking the inclusion of one axion sideband.
    }
    \label{fig:sigma_x}
\end{figure}
Fig.~\ref{fig:sigma_x} presents the noisy simulation results for the spin polarization expectation value $\expval{x}$. After application of the filtering procedure (see Fig.~\ref{fig:sigma_x}(a)), a well‐defined oscillatory component emerges at the theoretically predicted modulation frequency interval, exhibiting close quantitative agreement with the corresponding noiseless prediction. In Fig.~\ref{fig:sigma_x}(c), it demonstrates the PSD for noise only, which is higher at the low frequency but still non-zero and negligible at the desired signal frequency.

As shown in Fig.~\ref{fig:sigma_x}(d), the normalized cumulative power exhibits discrete, step‐like increases not only at the qubit fundamental resonance frequency $f_0$ but also at the sideband frequencies $f_0 \pm f_{\text{mod}}$ instead (indicated by the vertical dashed lines). These inflection points coincide precisely with the theoretically predicted positions of the modulation‐induced sidebands corresponding to the peaks with vertical green lines in Fig.~\ref{fig:sigma_x}(b), demonstrating that a significant fraction of the desired signal energy is also carried by the axion‐driven sideband components rather than being only confined solely to the qubit itself. Consequently, the modulation power appears predominantly in these sidebands rather than at the un-modulated resonance.

\begin{figure}[htbp!]
    \centering
    \includegraphics[width=3.14in, keepaspectratio]{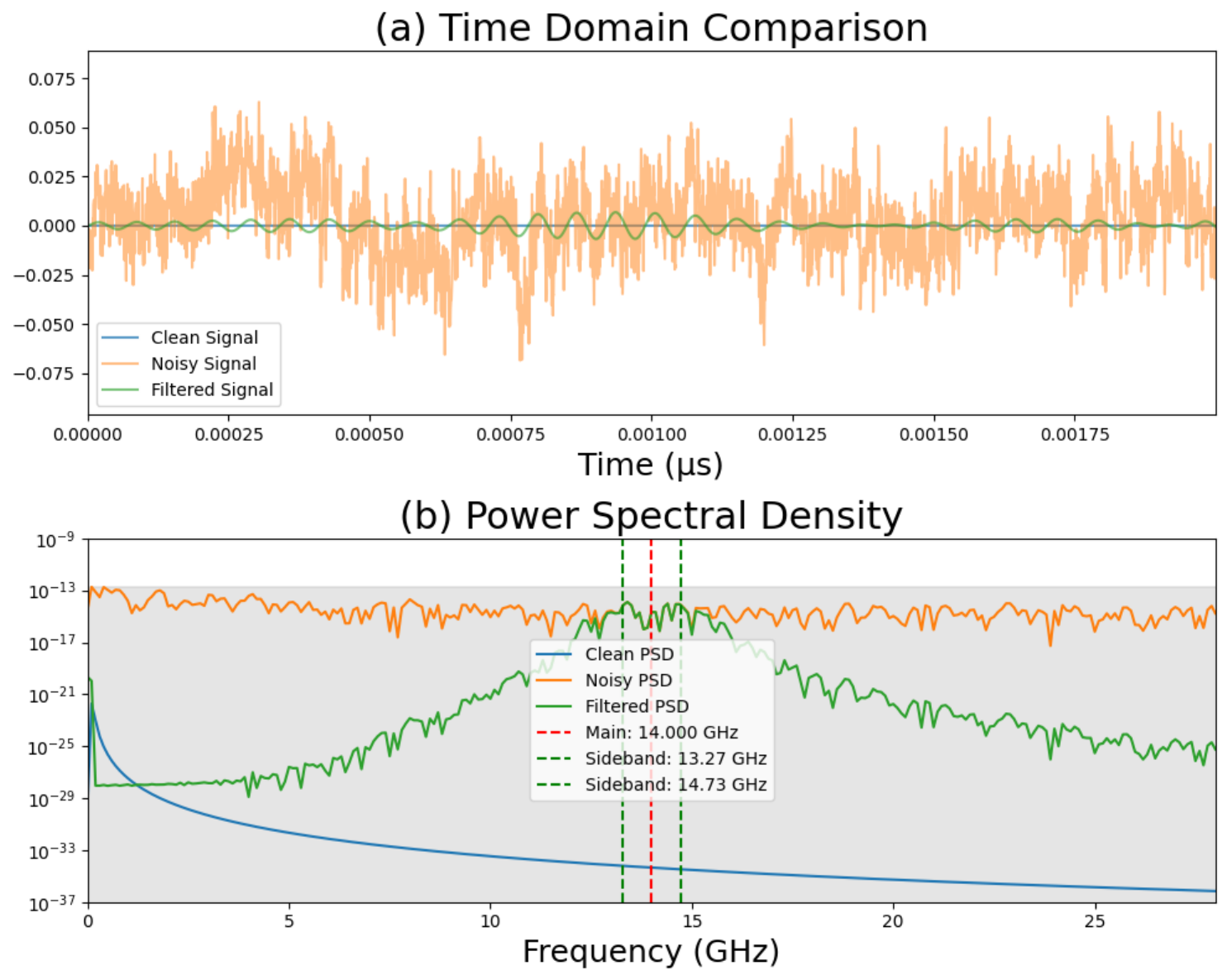}
    \caption{
        (a) Time-domain comparison of $\langle\sigma_z\rangle$: the ideal signal (blue) stays constant, 
        the raw noisy trace (orange) fluctuates around zero, and the filtered result (green) suppresses noise 
        without revealing coherent oscillations, consistent with pure phase evolution. 
        (b) Power spectral densities of the clean (blue), noisy (orange), and filtered (green) signals. 
        The carrier (red dashed) and sidebands (green dashed) are absent in the ideal case; 
        residual ripples in the filtered PSD arise from spectral leakage, imperfect filter roll-off, 
        and numerical noise rather than true longitudinal coupling.
    }
    \label{fig:sigma_z}
\end{figure}

The spin polarization $\expval{z}$ simulation result is shown in Fig.~\ref{fig:sigma_z}(a), manifests as a transient oscillation in the filtered trace (green) at the filter’s center frequency. To eliminate these edge-transient oscillations, one can use a zero–phase forward–backwards filter (for example, by using the \texttt{filtfilt} function), which applies the filter in both directions and cancels out the phase delay and associated ringing even after noise filtering.
Note that in z-axis signal processing, applying a causal band‐pass filter to a signal containing abrupt transitions or broadband noise will convolve the data with the filter’s oscillatory impulse response, producing decaying sinusoidal ringing artifacts-even when the actual underlying signal is zero, since the $\expval{\sigma_z}$ should always be zero without the noise unlike the $\expval{\sigma_x}$. And in this case, the signal level is below the noise floor, which means there is no axion influence.

Despite our focus was using the electron spin qubits in a quantum dot array as a prob to detect the axion-qubit couplings, we notice that there are alternative architectures such as using hole carriers in the valence band as the qubit platform, for its longer relaxation time $T_1$, and dephasing time $T_2^*$ \cite{Wang2021}.
Hole-based qubits benefit from improved dephasing times because the $p$-orbital symmetry of their wavefunctions excludes contact hyperfine interactions \cite{Jehl2016}.
Prevalent hole spin qubit structure concentrate on group IV materials like germanium and silicon, whose valence band is described by the Luttinger-Kohn Hamiltonian \cite{Yinan2023}.
The confinement potentials takes the same form as we seen in the electronic spin qubits, but the total Zeeman Hamiltonian and strain Bir-Pikus Hamiltonian for strains now include heavy-hole band, light-hole band, and the split-off band \cite{Luttinger1955}.
Due to the complex valance band structure, which possess the $p$-orbit symmetries, the modulation of the $g$-factors is much larger.
For example, in silicon hole spin qubits, experiments have reported up to 500$\%$ $g$-factor modulation owing to the combined effects of strain and spin–orbit coupling.
Meanwhile, the spin–orbit coupling mechanisms are also more intricate: in addition to a term analogous to the linear Rashba spin–orbit coupling seen in electron spin qubits, there exists a cubic Rashba spin–orbit coupling term permitted by symmetry.
The linear spin-orbit couplings in hole systems read:
\begin{equation}
    H_{\text{SO,1}} = \alpha_{\text{R}_1}(k_x \sigma_y + k_y \sigma_x) + \alpha_{\text{D}}(k_x\sigma_x+k_y\sigma_y) \,.
\end{equation}
The first term is the linear Rashba spin-orbit coupling term, and the second term is the Dresselhaus-like spin-orbit coupling term.
The latter arises from the combined effects of structure inversion asymmetry (induced by the gate electric field) and bulk inversion asymmetry (resulting from strain profiles within the material).
The cubic Rashba spin-orbit couplings will read:
\begin{equation}
\begin{aligned}
    H_{\text{SO,3}} =& \alpha_{\text{R}_2}(k_+^3 \sigma_- - k_-^3 \sigma_+) \\
    &+ \alpha_{\text{R}_3}(k_+k_-k_+ \sigma_- - k_-k_+k_- \sigma_+) \,,
\end{aligned}
\end{equation}
where $k_\pm=k_x\pm iky$, $\sigma_\pm=\sigma_x \pm i \sigma_y$.
This term reflects the inversion asymmetry induced by the gate electric field.
As a result, hole systems are expected to allow broader tunability of the qubit Larmor frequency through both gate electric field and strain engineering, which may extend the accessible search range of our method.
In addition, strong spin–orbit coupling and magnetic field effects arising from gauge terms can tilt the spin quantization axis.
Therefore, the spin dynamics in hole systems are expected to be more complex, potentially leading to distinct modulation patterns in the spin response.

Post-filter spectral analysis demonstrates that the dynamic SNR near the sideband can reach up to $20$dB, underscoring the efficacy of our method for $5\sigma$ high-sensitivity axion searches. Notwithstanding the comprehensive noise model (cf.\ Table~\ref{tab:noise-config}), the observed SNR may be artificially elevated due to several factors. First, the parametric noise model may not fully capture all real-world noise sources, such as ambient temperature fluctuations, radio-frequency leakage, and impedance mismatches at critical device interfaces, which can result in residual technical noise.
Second, non-idealities in the actual passband response of the filtering chain and lock-in detection stages can result in overestimated noise suppression, and also the non-ideality of the measurement apparatus.
Finally, calibration uncertainties in amplifier gain and systematic biases introduced by the choice of window functions and overlap strategies during Fourier analysis can further inflate the apparent dynamic SNR.

Compared to the previous work in superconducting systems, spin qubits offer excellent tunability and control via both electric and magnetic fields, making them a versatile platform for scanning a wide range of resonant frequencies that could coincide with potential axion signals. Their relatively long coherence times, compared to other solid-state qubits, enhance sensitivity to weak axion-induced perturbations. Moreover, the compatibility of semiconductor spin qubits with existing CMOS fabrication techniques enables large-scale integration, paving the way for the development of array-based sensor networks for high-sensitivity axion detection.

However, we emphasize that although our method can be adapted to various spin qubit platforms, the inherently weak interaction between axions and conventional matter still requires ultra-low-noise environments and cryogenic operating temperatures, which increase experimental complexity. Residual noise sources, such as hyperfine interactions, charge fluctuations, and magnetic field inhomogeneities, can further limit coherence times and reduce the SNR.
Moreover, accurately reading out these qubits requires advanced measurement and amplification techniques, which introduce additional experimental overhead and potential noise pathways that may be beyond the scope of this work.
\section{Conclusion and outlook}
\label{S4: Conclusion and outlook}
In this work, we introduce a novel approach to axion detection using semiconductor quantum dot spin qubits.
Relying on the axion–electron coupling mechanism, our scheme employs spin qubits as quantum sensors to transduce axion-induced electromagnetic perturbations into measurable spin dynamics along the $x$ and $y$ axes.
We construct a theoretical framework that describes this interaction and verify it through comprehensive numerical simulations, which show excellent agreement with our analytical predictions.
To isolate the axion-modulated signal from other irrelevant frequencies, we design a bespoke filter whose performance is evaluated under a noise model parameterized by experimental data, thereby faithfully reproducing realistic laboratory conditions.
%
\end{document}